\documentclass{article}

\usepackage{PRIMEarxiv}
\usepackage{amsmath} 
\usepackage[utf8]{inputenc} 
\usepackage[T1]{fontenc}    
\usepackage{hyperref}       
\usepackage{url}            
\usepackage{booktabs}       
\usepackage{amsfonts}       
\usepackage{nicefrac}       
\usepackage{microtype}      
\usepackage{lipsum}
\usepackage{fancyhdr}       
\usepackage{graphicx}
\usepackage[table,xcdraw]{xcolor}
\usepackage{colortbl}
\usepackage{array}
\graphicspath{{media/}}     

\title{SCALE: Self-regulated Clustered federAted LEarning in a Homogeneous Environment
}

\author{
       Sai Puppala, Ismail Hossain, Md Jahangir Alam, Sajedul Talukder \\
      School of Computing \\
      Southern Illinois University Carbondale, IL, USA, 62901\\
      \texttt{\{saimaniteja.puppala, ismail.hossain, mdjahangir.alam, sajedul.talukder\}@siu.edu} 
      \\
      \And
      Zahidur Talukder\\
      The University of Texas at Arlington, USA\\
    \texttt{zahidurrahim.talukder@mavs.uta.edu} 
    \\
      \And
      Syed Bahauddin\\
      University of Illinois Urbana-Champaign, IL, USA\\
      \texttt{alams@illinois.edu} 
}

\begin{document}
\maketitle

\begin{abstract}
Federated Learning (FL) has emerged as a transformative approach for enabling distributed machine learning while preserving user privacy, yet it faces challenges like communication inefficiencies and reliance on centralized infrastructures, leading to increased latency and costs. This paper presents a novel FL methodology that overcomes these limitations by eliminating the dependency on edge servers, employing a server-assisted Proximity Evaluation for dynamic cluster formation based on data similarity, performance indices, and geographical proximity. Our integrated approach enhances operational efficiency and scalability through a Hybrid Decentralized Aggregation Protocol, which merges local model training with peer-to-peer weight exchange and a centralized final aggregation managed by a dynamically elected driver node, significantly curtailing global communication overhead. Additionally, the methodology includes Decentralized Driver Selection, Check-pointing to reduce network traffic, and a Health Status Verification Mechanism for system robustness. Validated using the breast cancer dataset, our architecture not only demonstrates a nearly tenfold reduction in communication overhead but also shows remarkable improvements in reducing training latency and energy consumption while maintaining high learning performance, offering a scalable, efficient, and privacy-preserving solution for the future of federated learning ecosystems.
\end{abstract}

\section{Introduction}
\label{sec:intro}
Collaborative model training in FL is achieved by breaking down the training process into local training and model aggregation stages~\cite{bonawitz2019towards}. Each data owner conducts local training on its specific data partition and shares only intermediate results, such as gradients, for model aggregation. This communication can take place either at a centralized server or directly among data owners. The version of federated learning that employs a central server for coordinating model aggregation is referred to as centralized FL. Centralized FL aims to connect thousands of diverse, distributed devices into a cloud-based centralized server, where devices' local models are directly transmitted to the cloud server. This traditional approach is plagued by issues such as limited wireless resources and long transmission distances, leading to unpredictable and unreliable communication that can severely impair training efficiency and model accuracy. 

Addressing this critical bottleneck of communication inefficiency, the Hierarchical Federated Learning (HFL) framework has emerged as a promising solution. By decentralizing the process, HFL employs a two-tier aggregation process where local parameter values are initially aggregated at edge servers, such as base stations, before undergoing global aggregation at the central server~\cite{xu2021adaptive}. This goal of significantly reducing communication overhead with the cloud lead to a client-edge-cloud hierarchical Federated Learning (FL) system, aiming to combine the reduced data access of cloud servers with the rapid updates of edge servers~\cite{liu2020client}. To enhance resource allocation and incentive design at the edge servers, decentralized edge intelligence has been proposed~\cite{lim2021decentralized} that employs a deep learning based auction mechanism. However, a significant limitation of existing HFL approaches is the dependency on a specialized, and often costly, edge server infrastructure that necessitates preliminary setup, rendering it impractical for dynamic edge environments. This underscores the need for innovative HFL strategies that are both cost-effective and flexible, capable of adapting to the ever-changing dynamics of edge computing environments without sacrificing efficiency or scalability.  


This paper introduces a self-regulated clustered federated learning architecture, known as SCALE that transcends the conventional reliance on edge servers for clustering, by employing a Hybrid Decentralized Aggregation Protocol (HDAP) based on dynamic driver node selection. Our approach offers a direct, server-assisted mechanism for optimizing cluster formation that leverages Proximity Evaluation, integrating data similarity, performance indices, and geographical proximity to dynamically cluster edge devices. The process starts by prioritizing initial data preparation, where key components like feature variance and performance indices are computed at the client node, then securely transmitted to the global server for cluster formation. This ensures that clusters are intelligently formed, reflecting a balance between data relevance, device capability, and geographical proximity to enhance overall communication and computational efficiency. This strategy not only mitigates privacy and efficiency concerns but also significantly reduces the dependency on costly edge infrastructure, offering a scalable and flexible solution adaptable to the fluid nature of edge computing environments.

The main contributions of this paper are summarized as follows:

\begin{itemize}
    \item We introduce a novel Federated Learning (FL) architecture that eliminates the need for traditional edge server infrastructure, leveraging a server-assisted Proximity Evaluation mechanism for dynamic cluster formation. This approach significantly mitigates privacy and efficiency concerns associated with conventional FL systems.
    \item Our methodology integrates data similarity, performance indices, and geographical proximity to optimize cluster configurations, enhancing operational efficiency and scalability within the FL process.
    \item We propose a Hybrid Decentralized Aggregation Protocol that synergizes local model training with peer-to-peer weight exchange and a centralized final aggregation phase managed by a dynamically elected driver node, effectively reducing global communication overhead.
    \item The paper incorporates a Decentralized Driver Selection mechanism, a Check-pointing strategy to minimize network traffic, and a Health Status Verification Mechanism to ensure the robustness and reliability of the FL system.
    \item Through extensive simulations using the breast cancer dataset, we demonstrate superior performance of our architecture in reducing communication overhead, training latency, and energy consumption, while maintaining high learning performance compared to traditional FL systems.
\end{itemize}

The remainder of this paper is structured as follows: Section II delves into the background and related work, highlighting the evolution of FL frameworks and the pivotal role of clustering mechanisms. Section III presents our proposed methodology in detail, from initial data preparation to the final cluster formation and driver node selection. Section IV evaluates our approach through experimental validation, showcasing its advantages over conventional models. Finally, Section V concludes the paper with a discussion on the implications of our findings and directions for future research.

\begin{figure*}
    \centering
    \includegraphics[width=14cm]{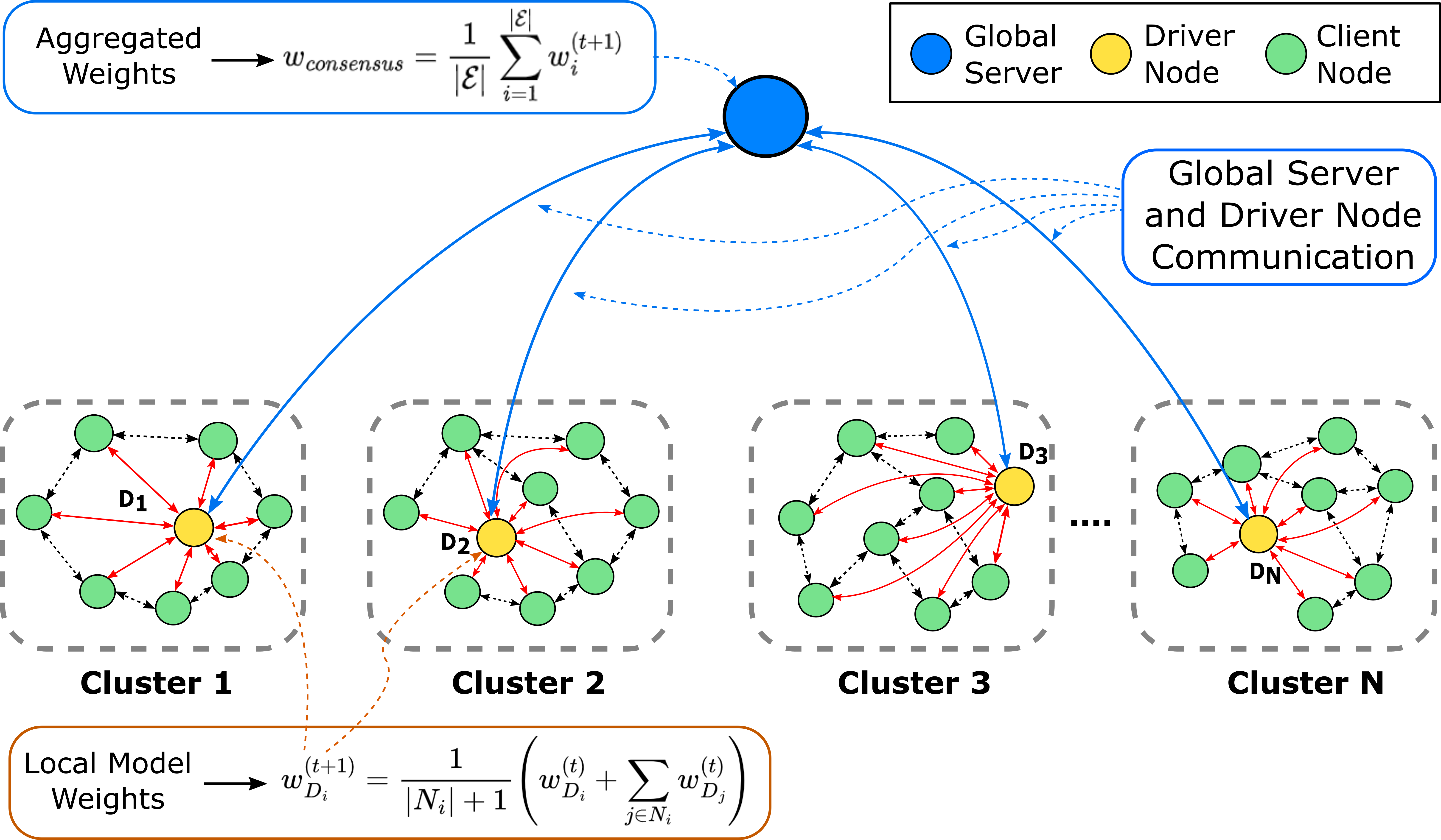}
    \caption{The system architecture elucidates various components in the SCALE approach, with a key focus on driver nodes, client nodes, and communication with the global server. Additionally, it provides a concise overview of the distinctive features of the global server. The significance of local aggregations and checkpoints has been underscored for each driver node, highlighting their essential role in the primary functionality of clusters.}
    \label{fig:enter-label}
\end{figure*}

\section{Related Works}

Since the inception of federated learning \cite{yang2019federated}, a significant objective has been to reduce the communication costs associated with Federated Learning \cite{konevcny2016federated}. However, as pointed out by Li et al. \cite{li2020federated}, there are limited existing techniques that alter the approach to user selection. Hartmann \cite{hartmann2018federated} proposes stratification based on contextual information about users, while Nishio and Yonetani \cite{nishio2019client} group users based on hardware characteristics. Author zheng in his paper talks about the uplink and downlink during communication between global server and client nodes \cite{zheng2020design}. We might see communications as not a big problem, provided if we have a enough bandwidth, but over communications will impact battery life. 
Communication overload on global server and cost implications are commonly addressed open problems in federated learning architecture. on the other side, author Aroju talks about the energy aware systems in federated learning \cite{arouj2022towards}.

In a recent survey conducted by Zihao, efforts to enhance communication efficiency were explored \cite{zhao2023towards}. Additionally, numerous studies have been undertaken in the past to achieve swift communication establishment and low latency between global server and client nodes, aiming for expedited data processing \cite{almanifi2023communication}. Research advancements in addressing communication bottlenecks in federated learning include the exploration of communication-efficient algorithms that minimize information exchange between the central server and nodes, such as federated optimization algorithms emphasizing local updates \cite{shahid2021communication}. Quantization techniques compress model updates before transmission, reducing exchanged information size without significant performance loss \cite{alistarh2017qsgd}. Asynchronous federated learning allows independent node updates, reducing dependency on synchronized communication and enhancing scalability \cite{huba2022papaya}. Adaptive strategies for federated learning can be found in some of few research articles \cite{puppala2022towards,talukder2022federated,hossain2023collaborative,talukder2022novel}. These advancements collectively contribute to more efficient and scalable federated learning systems by mitigating the communication challenges inherent in decentralized and distributed settings.




\section{Methodology}

\subsection{Initial Data Preparation}
To accurately identify the driver node within a federated learning framework, we introduce an advanced initial data preparation method. This method involves computing several critical components at the client node, which are then encrypted and transmitted to the global server. These components include feature variance, and performance indices for edge devices. We provide a detailed mathematical and technical explanation for each component below.

\subsubsection{Feature Variance}
Feature variance calculation is pivotal for grouping similar datasets, thereby enhancing the federated averaging process's efficiency.

\noindent
\textbf{Method 1: Alphabetical Schema-Based Scoring}
Feature variance is computed at the client node by analyzing the metadata of input datasets, focusing on columns and their schema. To ensure consistent scoring for identical attributes, columns are arranged in alphabetical order. This ordering is crucial to avoid discrepancies in feature scoring.

Given a feature attribute represented by a string $a_7,a_6,\ldots,a_1,a_0$, the formula for calculating its score is as follows:
\begin{equation}
    \begin{split}
        \text{Score} = & a_7 \cdot 35^6 + a_6 \cdot 35^5 + a_5 \cdot 35^4 \\
        & + a_4 \cdot 35^3 + a_3 \cdot 35^2 + a_2 \cdot 35^1 + a_1 \cdot 35^0
    \end{split}
\end{equation}

where each character in the attribute name is assigned a numeric value based on its position in the English alphabet (A=0, B=1, ..., Z=25).

\noindent
\textbf{Method 2: Combined Metadata Features}
To further refine the feature variance score calculations, we adopt a method that utilizes combined metadata features. This approach incorporates the alphabetical order of columns and their data types. The combined metadata score, $M$, is calculated using a weighted sum of sorted columns, $C_{\text{sorted}}$, and their corresponding data types, $C_{\text{type}}$:
\begin{equation}
    M = w_{\text{sorted}} \cdot C_{\text{sorted}} + w_{\text{type}} \cdot C_{\text{type}}
\end{equation}
Weights $w_{\text{sorted}}$ and $w_{\text{type}}$ reflect the relative importance of each aspect in the combined metadata score.

\subsubsection{Performance Index for Edge Devices}
The Performance Index (P.I.) is crucial for evaluating an edge device's suitability as a driver node within a federated learning network. Given the unique constraints and capabilities of edge devices, we derive the P.I. from metrics such as energy efficiency, latency, network bandwidth, and concurrency support.

\noindent
\textbf{Method 1: Compute Ability Score for Edge Devices}
The Compute Ability Score serves as a foundational metric for the global server to cluster edge devices in a federated learning environment. By assessing computational power, energy efficiency, latency, network stability, and multitasking capabilities, this score enables the global server to categorize devices based on their performance and operational efficiency. This clustering facilitates optimized task allocation, ensuring devices are grouped and utilized according to their strengths and capabilities, enhancing the overall efficiency and effectiveness of federated learning processes.

The formula for calculating the Compute Ability Score integrates these attributes, each weighted according to its impact on the device's overall performance, thus providing a nuanced evaluation metric:

\begin{equation}
    x' = a + \frac{(x - \min(x))(b - a)}{\max(x) - \min(x)}
\end{equation}

These values are scaled to a uniform range via transformation, facilitating a balanced evaluation across diverse hardware specifications and operational conditions. Subsequently, the Compute Ability Score for an edge device is calculated as a weighted sum of these scaled values, integrating the various metrics into a comprehensive performance index:

\begin{equation}
    P.I_{\text{compute}} = w_{1} \cdot C_{p} + w_{2} \cdot E_{e} + w_{3} \cdot L + w_{4} \cdot N_{b} + w_{5} \cdot C_{l}
\end{equation}

Here, $C_{p}$ represents Computational Power, $E_{e}$ for Energy Efficiency, $L$ indicates Latency, $N_{b}$ measures Network Bandwidth, and $C_{l}$ reflects Concurrency Level. Each metric, scaled and assigned a weight ($w_{i}$), quantifies each device's operational efficiency and suitability to facilitate clustering.

\noindent
\textbf{Method 2: Operational Efficiency Score}
The Operational Efficiency Score assesses edge devices' performance in a federated learning setting by examining CPU utilization, energy consumption, network efficiency, and energy efficiency. This comprehensive metric evaluates how effectively a device manages its computational resources (CPU utilization), its power usage (energy consumption and energy efficiency), and its capacity to maintain stable communication (network efficiency). Leveraging this information, the global server can effectively cluster devices, aligning them based on their operational efficiency and resource optimization capabilities through a composite measure:
\begin{align}
    \psi &= \frac{1}{\text{CPU Utilization} \cdot w_1} + \frac{1}{\text{Energy Consumption} \cdot w_2} \nonumber \\
         &\quad + \frac{1}{\text{Network Efficiency} \cdot w_3} + \frac{1}{\text{Energy Efficiency} \cdot w_4}
\end{align}
\begin{equation}
    Local\ P.I(\alpha) = \frac{1}{\psi/4}
\end{equation}

Considering the varied nature of edge devices, a logarithmic transformation is applied to the performance index scores to ensure scalability and manageability before transmission to the global server:
\begin{equation}
    Local_{\log} P.I = \log_e(\alpha)
\end{equation}
This strategic approach enables the global server to allocate clustering more efficiently, ensuring that devices are utilized in a manner that enhances the network's overall performance and sustainability.

\subsection{Global Server-Assisted Parallel Integration for Cluster Formation}

To optimize the FL process, we propose a global server-assisted mechanism for forming clusters based on parallel integration of data similarity and geographical proximity. Edge devices submit their dataset summaries, performance indices, and geographical locations to the global server. This approach leverages a multi-dimensional strategy for cluster formation, ensuring that the resulting groups $\mathcal{C}$ are optimized not only based on computational capabilities and operational efficiency but also on the spatial distribution of nodes. By integrating these key factors, the server aims to enhance the overall effectiveness and efficiency of the federated learning process. 

Optimizing the FL process necessitates efficient cluster formation, as elaborated in Algorithm 2. The global server synthesizes data similarity $\mathcal{DS}$, performance index $\mathcal{PI}$, and geographical proximity $\mathcal{GP}$ to form optimized clusters $\mathcal{C}$. The integration process aims to minimize intra-cluster variance while maximizing inter-cluster distances, mathematically optimizing cluster homogeneity and geographical efficiency. The server then assigns each device to an optimized cluster that balances the relevance of data and communication efficiency.














\subsubsection{Proximity Evaluation}
This process involves assessing the geographical closeness of nodes to facilitate efficient data communication and computational collaboration among devices. By leveraging geographical coordinates, the global server can group devices that are physically closer, thereby reducing latency and improving the overall speed and reliability of the federated learning tasks. Additionally, proximity evaluation helps in minimizing network congestion and optimizing bandwidth usage, which are essential for the scalability and performance of federated learning systems.

\noindent
For an alternative proximity evaluation, the Equirectangular Approximation calculates distances between nodes with the formula:
\begin{equation}
    \text{distance} = R \cdot \sqrt{(\Delta\phi)^2 + (\cos(\frac{\phi_1 + \phi_2}{2}) \cdot \Delta\lambda)^2},
\end{equation}
where \(R\) is the Earth's radius, \(\Delta\phi\) and \(\Delta\lambda\) are the differences in latitude and longitude, and \(\phi_1\) and \(\phi_2\) represent the latitudes of the two points. This method offers a simpler calculation for distances, aiding in the efficient clustering of nodes based on geographic proximity.

\subsection{Hybrid Decentralized Aggregation Protocol for Federated Learning}

The FL framework is further refined through a hybrid decentralized aggregation protocol, which combines local model training and peer-to-peer weight exchange with centralized final aggregation under a dynamically selected driver. This protocol minimizes global communication overhead and maximizes the efficiency of model updates aggregation. The selection of a driver node is crucial for coordinating the final aggregation and ensuring the distributed learning process's cohesion and progression.

The Hybrid Decentralized Aggregation Protocol, introduces an innovative approach to model aggregation within the FL framework. It synergizes local model training and peer-to-peer weight exchange with a centralized final aggregation phase conducted by a dynamically selected driver. This hybrid protocol is designed to minimize global communication overhead while maximizing the efficiency and efficacy of the aggregation process. The mathematical formulation of the protocol's operation is as follows:\\

\textbf{Local Model Training and Weight Exchange:}
\begin{itemize}
    \item Each edge device $e_i \in \mathcal{E}$ updates its model weights $w_{i}^{(t)}$ through local training on its dataset $D_i$.
    \item Subsequently, $e_i$ engages in weight exchange with a selected subset of peers $N_i$, aggregating received weights to update $w_{i}^{(t+1)}$:
    \begin{equation}
    w_{i}^{(t+1)} = \frac{1}{|N_i| + 1} \left( w_{i}^{(t)} + \sum_{j \in N_i} w_{j}^{(t)} \right)
    \end{equation}
\end{itemize}

\textbf{Centralized Final Aggregation by Driver:}
\begin{itemize}
    \item A driver $L$ is elected through the decentralized driver selection mechanism (Algorithm 4), coordinating the final aggregation of model weights across the cluster.
    \item The driver $L$ computes the final aggregated model weights $w_{consensus}$ by averaging the updated weights from all devices in the cluster:
    \begin{equation}
    w_{consensus} = \frac{1}{|\mathcal{E}|} \sum_{i=1}^{|\mathcal{E}|} w_{i}^{(t+1)}
    \end{equation}
\end{itemize}

This protocol effectively leverages the distributed computing power of edge devices for local computations and peer-to-peer communications, reducing the reliance on the global server for frequent data exchanges. The inclusion of a centralized aggregation phase under a selected driver ensures coherent and efficient model updates, enhancing the overall learning process.






    
    
    





\subsection{Decentralized Driver Selection}

To maintain the system's operational continuity, we introduce a decentralized driver selection mechanism that takes place after decentralized weight exchange and averaging. This is also triggered upon the failure of a current driver node, alongside a health status verification mechanism to monitor the vitality of the communication channels. Several critical criteria ensure the elected driver optimizes the aggregation phase's efficiency and effectiveness: Computational Capacity, Network Connectivity and Bandwidth, Battery Life or Energy Resources, Reliability and Availability, Data Representativeness, and Security and Trustworthiness.

Computational Capacity is essential for processing and aggregating data swiftly, while Network Connectivity and Bandwidth ensure fast and reliable communication with other devices. Battery Life or Energy Resources guarantee the device can sustain the increased workload without interruption. Reliability and Availability reflect a device's historical uptime, indicating its dependability throughout the learning process. Data Representativeness is crucial for generating a global model that accurately reflects the collective dataset, and Security and Trustworthiness ensure the integrity of the data and the aggregation process, safeguarding against potential breaches or compromises. Together, these criteria form a comprehensive framework for selecting a driver capable of efficiently managing the complexities and demands of federated learning in an edge computing environment.

The Decentralized Driver Selection mechanism is presented in Algorithm 4. Upon a leadership vacuum, a new driver is elected based on predefined criteria $\mathcal{P} = \{p_1, p_2, \ldots, p_l\}$, incorporating factors like computational capacity and network stability. The election process is mathematically represented as:

\begin{equation}
L = \underset{e_i \in \mathcal{E}}{\arg\max} \left( \sum_{j=1}^{l} \omega_j \cdot p_{j,i} \right)
\end{equation}

where $L$ designates the new driver, $p_{j,i}$ indicates the $j$th criterion for device $e_i$, and $\omega_j$ is the weight assigned to the $j$th criterion. This formula ensures a weighted and consensus-based driver election, enhancing the democratic essence of the selection process.

\section{Experiment}
\label{sec:Experiment}
We initiated our experiment by utilizing the SCALE architecture with the breast cancer data-set, distributed among 100 client nodes. These data-sets were shared across all client nodes in both identical and non-identical ways. We evaluated our approach in two distinct federated learning scenarios. Initially, we employed the breast cancer data-set within the traditional federated learning framework, and subsequently, we compared the results with the SCALE approach.

\begin{figure}[h]
\centering
\includegraphics[width=8cm, height=9cm]{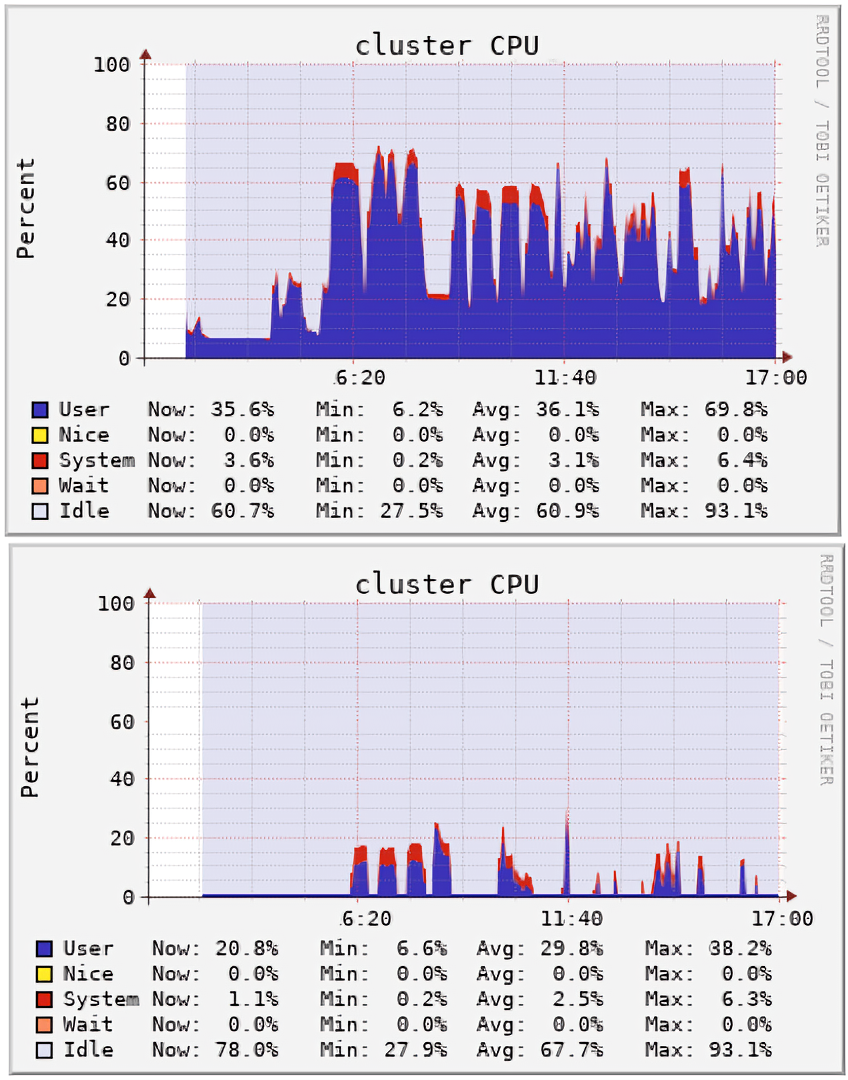}
\caption{The comparison metrics between different scales of traditional federated learning and the SCALE approach are under scrutiny. These statistical measures are derived at the conclusion of randomly selected epoch rounds. This analysis aims to provide a nuanced understanding of how the two approaches differ in performance across various evaluation criteria.} 
\label{fig:data_classifications}
\end{figure}

\subsection{Dataset}

The BreastHealth Dataset is a comprehensive and diverse collection of annotated mammogram images designed to support research and development in the field of medical imaging and breast cancer prediction, with a focus on healthcare applications. This dataset provides a rich source of mammogram images along with detailed annotations, making it an invaluable resource for various tasks such as breast cancer detection, classification, and risk assessment. We have gathered data from Breast Cancer Wisconsin \cite{misc_breast_cancer_wisconsin_(diagnostic)_17} and utilized 30 available features to forecast whether breast tissue is malignant or benign. 

In the context of local training and validation of accuracies for traditional federated learning and SCALE architecture, we have conducted our experiments using \textit{\textbf{Support Vector Classifier}} machine learning algorithm.

\subsection{Results}
Our empirical investigation systematically evaluates the SCALE approach against traditional federated learning frameworks, leveraging diverse performance metrics, communication efficiencies, and cost considerations. Our analysis, supported by comprehensive metrics across model performance, communication efficiency, and cost-effectiveness, reveals critical insights into the operational benefits of SCALE. Below, we delve deeper into these aspects, referencing the pertinent results for a thorough examination.

\subsubsection{Model Performance Metrics}
The quantitative analysis of model performance, as mentioned in Table \ref{tab:scale-communication-metrics}, highlights the nuanced improvements facilitated by SCALE. Initially, both methodologies display comparable accuracies. However, as training progresses, SCALE demonstrates a subtle yet consistent enhancement in model performance metrics, including F1 scores, precision, recall, and ROC AUC values. This incremental improvement underscores the efficacy of SCALE's distributed learning mechanism, which not only accommodates but leverages the diversity of data across nodes to refine model accuracy effectively.

\subsubsection{Communication Metrics}
A critical examination of the communication overhead, unveils SCALE's strategic reduction of global server updates. The stark reduction from 2850 updates in the federated learning paradigm to a mere 235 under SCALE underscores a significant optimization in data transmission. This efficiency is pivotal in mitigating network congestion and minimizing latency, thereby enhancing the overall scalability of the federated learning system. Such an approach is not merely a technical improvement but a strategic reorientation towards more sustainable and efficient federated learning operations.

\subsubsection{Processing Latency}
Notably, the introduction of checkpointing mechanisms yields a dramatic reduction in latency, optimizing the aggregation process at the global server level. This methodological refinement enhances the responsiveness of the learning system, facilitating quicker model updates and enabling more agile adaptations to evolving data landscapes. The implications of such efficiency extend beyond mere time savings, suggesting a more dynamic and responsive federated learning framework.

\begin{table}[]
\centering
\begin{tabular}{|l|c|c|c|c|c|c|}
\hline
\multicolumn{3}{|c|}{\cellcolor[HTML]{FFCE93}\textbf{Global Communication Stats}} & \multicolumn{2}{c|}{\cellcolor[HTML]{FFCE93}\textbf{Fed Learning}} & \multicolumn{2}{c|}{\cellcolor[HTML]{FFCE93}\textbf{SCALE}} \\ \hline
\multicolumn{1}{|c|}{\cellcolor[HTML]{EFEFEF}\textbf{Runs}} & \multicolumn{1}{c|}{\cellcolor[HTML]{EFEFEF}\textbf{Nodes}} & \multicolumn{1}{c|}{\cellcolor[HTML]{EFEFEF}\textbf{Rounds}} & \multicolumn{1}{c|}{\cellcolor[HTML]{EFEFEF}\textbf{Updates}} & \cellcolor[HTML]{EFEFEF}\textbf{Acc} & \multicolumn{1}{c|}{\cellcolor[HTML]{EFEFEF}\textbf{Updates}} & \cellcolor[HTML]{EFEFEF}\textbf{Acc} \\ \hline
\textit{Cluster 1} & 9 & 30 & 270 & 0.93 & 29 & 0.91 \\ \hline
\textit{Cluster 2} & 9 & 30 & 270 & 0.88 & 29 & 0.86 \\ \hline
\textit{Cluster 3} & 11 & 30 & 330 & 0.81 & 30 & 0.85 \\ \hline
\textit{Cluster 4} & 10 & 30 & 300 & 0.90 & 20 & 0.89 \\ \hline
\textit{Cluster 5} & 10 & 30 & 300 & 0.86 & 17 & 0.86 \\ \hline
\textit{Cluster 6} & 10 & 30 & 300 & 0.82 & 28 & 0.85 \\ \hline
\textit{Cluster 7} & 12 & 30 & 360 & 0.91 & 7 & 0.86 \\ \hline
\textit{Cluster 8} & 9 & 30 & 270 & 0.81 & 21 & 0.78 \\ \hline
\textit{Cluster 9} & 8 & 30 & 240 & 0.84 & 30 & 0.89 \\ \hline
\textit{Cluster 10} & 12 & 30 & 210 & 0.83 & 24 & 0.86 \\ \hline
\textit{\textbf{Total}} & \textbf{100} & \textbf{30} & \textbf{2850} & \textbf{0.85} & \textbf{235} & \textbf{0.86} \\ \hline
\end{tabular}
\caption{Comparison metrics at cluster level of both SCALE and traditional approach. Each row represents the total number of communications sent for model updates and the average accuracies at global server level from edge nodes.}
\label{tab:scale-communication-metrics}
\end{table}

\subsubsection{Cost Implications}
By significantly reducing the computational demands on the global server and optimizing the efficiency of data transmission, SCALE offers a cost-effective solution for federated learning deployments. This cost efficiency is particularly salient in cloud-based implementations, where computational resources come at a premium. The SCALE approach, therefore, not only enhances the technical performance of federated learning systems but also presents a financially sustainable model for large-scale deployments.

\section{Conclusion}
In conclusion, this paper presents an innovative Federated Learning (FL) methodology that overcomes traditional challenges such as privacy concerns, high latency, and increased costs in distributed machine learning by eliminating edge server dependency and implementing a server-assisted Proximity Evaluation for dynamic cluster formation. This is based on data similarity, performance indices, and geographical proximity. Our approach includes a Hybrid Decentralized Aggregation Protocol that merges local model training with peer-to-peer weight exchange and centralized final aggregation, significantly reducing communication overhead. Additionally, Decentralized Driver Selection, Check-pointing mechanisms, and a Health Status Verification Mechanism enhance network efficiency, system robustness, and data privacy. Tested on a breast cancer dataset, our methodology drastically cuts communication overhead, improves training latency and energy efficiency, and maintains high learning performance, offering a scalable, efficient, and privacy-preserving framework for advancing federated learning.

\section{Acknowledgment}
This research was supported by NSF grant CNS-2153482.

\begin{thebibliography}{10}
\providecommand{\url}[1]{#1}
\csname url@samestyle\endcsname
\providecommand{\newblock}{\relax}
\providecommand{\bibinfo}[2]{#2}
\providecommand{\BIBentrySTDinterwordspacing}{\spaceskip=0pt\relax}
\providecommand{\BIBentryALTinterwordstretchfactor}{4}
\providecommand{\BIBentryALTinterwordspacing}{\spaceskip=\fontdimen2\font plus
\BIBentryALTinterwordstretchfactor\fontdimen3\font minus \fontdimen4\font\relax}
\providecommand{\BIBforeignlanguage}[2]{{%
\expandafter\ifx\csname l@#1\endcsname\relax
\typeout{** WARNING: IEEEtran.bst: No hyphenation pattern has been}%
\typeout{** loaded for the language `#1'. Using the pattern for}%
\typeout{** the default language instead.}%
\else
\language=\csname l@#1\endcsname
\fi
#2}}
\providecommand{\BIBdecl}{\relax}
\BIBdecl

\bibitem{bonawitz2019towards}
K.~Bonawitz, H.~Eichner, W.~Grieskamp, D.~Huba, A.~Ingerman, V.~Ivanov, C.~Kiddon, J.~Kone{\v{c}}n{\`y}, S.~Mazzocchi, B.~McMahan \emph{et~al.}, ``Towards federated learning at scale: System design,'' \emph{Proceedings of machine learning and systems}, vol.~1, pp. 374--388, 2019.

\bibitem{xu2021adaptive}
B.~Xu, W.~Xia, W.~Wen, P.~Liu, H.~Zhao, and H.~Zhu, ``Adaptive hierarchical federated learning over wireless networks,'' \emph{IEEE Transactions on Vehicular Technology}, vol.~71, no.~2, pp. 2070--2083, 2021.

\bibitem{liu2020client}
L.~Liu, J.~Zhang, S.~Song, and K.~B. Letaief, ``Client-edge-cloud hierarchical federated learning,'' in \emph{ICC 2020-2020 IEEE International Conference on Communications (ICC)}.\hskip 1em plus 0.5em minus 0.4em\relax IEEE, 2020, pp. 1--6.

\bibitem{lim2021decentralized}
W.~Y.~B. Lim, J.~S. Ng, Z.~Xiong, J.~Jin, Y.~Zhang, D.~Niyato, C.~Leung, and C.~Miao, ``Decentralized edge intelligence: A dynamic resource allocation framework for hierarchical federated learning,'' \emph{IEEE Transactions on Parallel and Distributed Systems}, vol.~33, no.~3, pp. 536--550, 2021.

\bibitem{yang2019federated}
Q.~Yang, Y.~Liu, T.~Chen, and Y.~Tong, ``Federated machine learning: Concept and applications,'' \emph{ACM Transactions on Intelligent Systems and Technology (TIST)}, vol.~10, no.~2, pp. 1--19, 2019.

\bibitem{konevcny2016federated}
J.~Kone{\v{c}}n{\`y}, H.~B. McMahan, F.~X. Yu, P.~Richt{\'a}rik, A.~T. Suresh, and D.~Bacon, ``Federated learning: Strategies for improving communication efficiency,'' \emph{arXiv preprint arXiv:1610.05492}, 2016.

\bibitem{li2020federated}
T.~Li, A.~K. Sahu, A.~Talwalkar, and V.~Smith, ``Federated learning: Challenges, methods, and future directions,'' \emph{IEEE signal processing magazine}, vol.~37, no.~3, pp. 50--60, 2020.

\bibitem{hartmann2018federated}
F.~Hartmann, ``Federated learning,'' \emph{Freie Universit{\"a}t Berlin}, 2018.

\bibitem{nishio2019client}
T.~Nishio and R.~Yonetani, ``Client selection for federated learning with heterogeneous resources in mobile edge,'' in \emph{ICC 2019-2019 IEEE international conference on communications (ICC)}.\hskip 1em plus 0.5em minus 0.4em\relax IEEE, 2019, pp. 1--7.

\bibitem{zheng2020design}
S.~Zheng, C.~Shen, and X.~Chen, ``Design and analysis of uplink and downlink communications for federated learning,'' \emph{IEEE Journal on Selected Areas in Communications}, vol.~39, no.~7, pp. 2150--2167, 2020.

\bibitem{arouj2022towards}
A.~Arouj and A.~M. Abdelmoniem, ``Towards energy-aware federated learning on battery-powered clients,'' in \emph{Proceedings of the 1st ACM Workshop on Data Privacy and Federated Learning Technologies for Mobile Edge Network}, 2022, pp. 7--12.

\bibitem{zhao2023towards}
Z.~Zhao, Y.~Mao, Y.~Liu, L.~Song, Y.~Ouyang, X.~Chen, and W.~Ding, ``Towards efficient communications in federated learning: A contemporary survey,'' \emph{Journal of the Franklin Institute}, 2023.

\bibitem{almanifi2023communication}
O.~R.~A. Almanifi, C.-O. Chow, M.-L. Tham, J.~H. Chuah, and J.~Kanesan, ``Communication and computation efficiency in federated learning: A survey,'' \emph{Internet of Things}, p. 100742, 2023.

\bibitem{shahid2021communication}
O.~Shahid, S.~Pouriyeh, R.~M. Parizi, Q.~Z. Sheng, G.~Srivastava, and L.~Zhao, ``Communication efficiency in federated learning: Achievements and challenges,'' \emph{arXiv preprint arXiv:2107.10996}, 2021.

\bibitem{alistarh2017qsgd}
D.~Alistarh, D.~Grubic, J.~Li, R.~Tomioka, and M.~Vojnovic, ``Qsgd: Communication-efficient sgd via gradient quantization and encoding,'' \emph{Advances in neural information processing systems}, vol.~30, 2017.

\bibitem{huba2022papaya}
D.~Huba, J.~Nguyen, K.~Malik, R.~Zhu, M.~Rabbat, A.~Yousefpour, C.-J. Wu, H.~Zhan, P.~Ustinov, H.~Srinivas \emph{et~al.}, ``Papaya: Practical, private, and scalable federated learning,'' \emph{Proceedings of Machine Learning and Systems}, vol.~4, pp. 814--832, 2022.

\bibitem{puppala2022towards}
S.~Puppala, I.~Hossain, and S.~Talukder, ``Towards federated learning based contraband detection within airport baggage x-rays,'' in \emph{2022 IEEE International Conference on Machine Learning and Applied Network Technologies (ICMLANT)}.\hskip 1em plus 0.5em minus 0.4em\relax IEEE, 2022, pp. 1--6.

\bibitem{talukder2022federated}
S.~Talukder, S.~Puppala, and I.~Hossain, ``Federated learning-based contraband detection within airport baggage x-rays,'' \emph{Journal of Computing Sciences in Colleges}, vol.~38, no.~3, pp. 218--218, 2022.

\bibitem{hossain2023collaborative}
I.~Hossain, S.~Puppala, and S.~Talukder, ``Collaborative differentially private federated learning framework for the prediction of diabetic retinopathy,'' in \emph{2023 IEEE 2nd International Conference on AI in Cybersecurity (ICAIC)}.\hskip 1em plus 0.5em minus 0.4em\relax IEEE, 2023, pp. 1--6.

\bibitem{talukder2022novel}
S.~Talukder, S.~Puppala, and I.~Hossain, ``A novel hierarchical federated learning with self-regulated decentralized clustering,'' \emph{Journal of Computing Sciences in Colleges}, vol.~38, no.~3, pp. 222--223, 2022.

\bibitem{misc_breast_cancer_wisconsin_(diagnostic)_17}
M.~O. S.~N. Wolberg, William and W.~Street, ``{Breast Cancer Wisconsin (Diagnostic)},'' UCI Machine Learning Repository, 1995, {DOI}: https://doi.org/10.24432/C5DW2B.

\end{thebibliography}

\end{document}